\documentclass[final,copyright]{eptcs}
\providecommand{\event}{MBT 2015} 
\usepackage{breakurl}             
\usepackage{graphicx}             
\usepackage{wrapfig}              
\usepackage{caption}              
\usepackage{array}                

\title{A Visual Formalism for Interacting Systems}
\author{Paul C. Jorgensen
\institute{School of Computing and Information Systems\\
Grand Valley State University\\
Allendale, Michigan USA}
\email{jorgensp@gvsu.edu}
}
\def\titlerunning{A Visual Formalism for Interacting Systems}
\def\authorrunning{P. Jorgensen}
\begin{document}
\maketitle

\begin{abstract}
Interacting systems are increasingly common—many examples pervade our 
everyday lives: automobiles, aircraft, defense systems, telephone 
switching systems, financial systems, national governments, and so on. 
Closer to computer science, embedded systems and Systems of Systems are 
further examples of interacting systems. Common to all of these is that 
some "whole" is made up of constituent parts, and these parts interact 
with each other. By design, these interactions are intentional, but it 
is the unintended interactions that are problematic. The Systems of 
Systems literature uses the terms "constituent systems" and 
"constituents" to refer to systems that interact with each other. That 
practice is followed here.

This paper presents a visual formalism, Swim Lane Event-Driven Petri 
Nets, that is proposed as a basis for Model-Based Testing (MBT) of 
interacting systems. In the absence of available tools, this model 
can only support the offline form of Model-Based Testing. 
\end{abstract}

\section{Existing Models for Interacting Systems}

To support offline MBT of interacting systems, a model must be 
capable of expressing the ways in which constituents interact. The 
best known, and most widely used, such model is the Statechart model 
\cite{Harel1988}. Statecharts have been incorporated into the Unified 
Modeling Language (UML) and further codified into Types, I, II, and 
III of UML Statecharts. The model presented in this paper has been 
shown to be formally equivalent \cite{DeVries2013} to all three types of 
UML statecharts.

Statecharts contain orthogonal regions, and these nicely represent 
distinct devices. As such, they can also be used to represent 
systems that interact with each other. The statechart broadcasting 
mechanism is the only vehicle for communicating interactions among 
the orthogonal devices/components. There is an elaborate language on 
transitions among blobs in an orthogonal region. Taken together, 
these notations result in a very "dense" model that is best 
understood by executing the statechart with an engine. There can be 
no doubt about the expressive power of statecharts. In a 
conversation \cite{iLogix1990} at a Grand Rapids (Michigan) avionics 
company, representatives of the i-Logix company related a success 
story in which exhaustive execution of a statechart model of a 
fully-deployed ballistic missile launch control system revealed a 
legitimate sequence of events that would launch a missile that was 
not known to the developing defense contractor.

Contemporaneously with Harel's work on statecharts, a North American 
industry group proposed the Extended Systems Modeling Language 
(ESML) which described how activities in a traditional data flow 
diagram could communicate with each other \cite{Bruyn1988}. The ESML 
prompts allowed an activity in a data flow diagram to Enable, 
Disable, Activate, Pause, Resume, Suspend, or Trigger another 
activity. Synonyms of many of these verbs are available in the 
statechart transition language.

\section{Comparing Statecharts with Swim Lane Event-Driven Petri Nets}

While they are clearly a powerful modeling technique, there are some 
problems with statecharts (as originally defined). Statecharts: 
\begin{itemize}
  \item Are best understood when executed by a customer using a statechart engine,
  \item Are a top-down model,
  \item Can only be composed under very limited circumstances 
  \cite{Harel1998}, and
  \item Are both rigorous and complex, a five-day training course is 
  recommended \cite{iLogix1991}.
\end{itemize}

These limitations are all answered by Swim Lane Event-Driven Petri 
Nets, specifically they are: 
\begin{itemize}
  \item intuitively clear, once the basic mechanism of Petri Net transition firing is understood,
  \item a bottom-up model. As such, they work well in agile developments.
  \item easily composed, particularly if the composition is accomplished in a database with well-designed queries.
  \item easy to learn
\end{itemize}

\section{Event-Driven Petri Nets and Swim Lane Petri Nets}

Basic Petri nets need two slight enhancements to become Event-Driven 
Petri Nets (EDPNs) \cite{Jorgensen2014}. The first enables them to 
express more closely event-driven systems, and the second deals with 
Petri net markings that express event quiescence, an important 
notion in object-oriented applications. Taken together, these 
extensions result in an effective, operational view of software 
requirements.

\newcommand{\jorgX}{
  {\sffamily \Large \textbackslash \hspace{-0.83em} /}
}

\begin{wrapfigure}{R}{5.7cm}
  \vspace{-15pt}
  \begin{center}
    \includegraphics{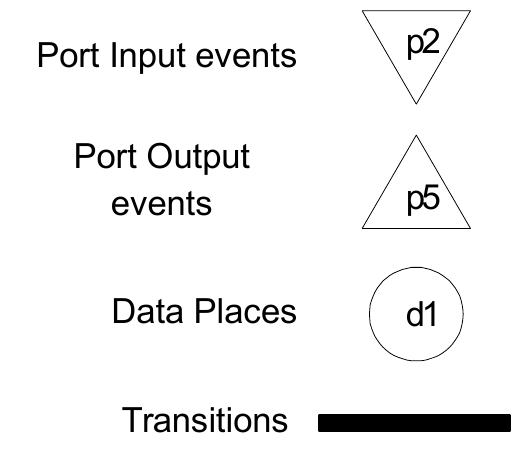}
  \end{center}
  \vspace{-10pt}
  \caption{Event-Driven\newline Petri Net elements}
  \label{fig1}
  \vspace{-50pt}
\end{wrapfigure}

\textbf{Definition}: An Event-Driven Petri Net (EDPN) is a 
tripartite-directed graph $(P, D, S, In, Out)$ composed of three 
sets of nodes, $P$, $D$, and $S$, and two mappings, $In$ and $Out$, 
where:
  \begin{itemize}
    \item $P$ is a set of port events
    \item $D$ is a set of data places
    \item $S$ is a set of transitions
    \item $In$ is a set of ordered pairs from $(P \cup D)$ 
      \hspace{-0.5em} \jorgX $S$ 
    \item $Out$ is a set of ordered pairs from $S$ \hspace{-0.3em} \jorgX 
      $(P \cup D)$ 
  \end{itemize}

The drawing conventions for EDPNs are in Figure \ref{fig1}. Other than 
explicit representation of discrete events, EDPNs are very similar 
to ordinary Petri nets. The other main difference is that an 
ordinary Petri net is a closed system, in which tokens and markings 
are determined only by transition firing. EDPNs support a concept of 
event quiescence, which is an extension to deadlock in an ordinary 
Petri net. Since events are (usually) from external devices, an EDPN 
more accurately represents an event-driven system (so EDPNs 
represent "open" systems.) Both ordinary and Event-Driven Petri Nets 
can be placed into UML-style "swim lanes." The most convenient 
interpretations of swim lanes is that they "contain" interacting 
constituent systems.

\section{Communication Primitives for Interacting Systems}

Integration testing for a single system is based on the assumption 
that the units being integrated have all been individually, and 
thoroughly, tested. The usual goals of integration testing are to 
find faults that are due to interfaces among the units, and as such, 
would not be revealed by unit testing. This clearly extends to 
testing interacting systems where we assume that the constituent 
systems are all thoroughly tested and function correctly. The goal 
of testing for interacting systems is to focus on the ways in which 
constituent systems communicate; here we present a set of 
communication primitives for that purpose. The first two are from 
ordinary Petri nets, followed by the communication primitives from 
the Extended Systems Modeling Language (ESML) \cite{Bruyn1988}. The ESML 
primitives are supplemented by three primitives that represent 
service requests among interacting systems.  Each primitive is 
briefly described and illustrated by accompanying figures next.

\begin{wrapfigure}{R}{5.7cm}
  \vspace{-35pt}
  \begin{center}
    \includegraphics{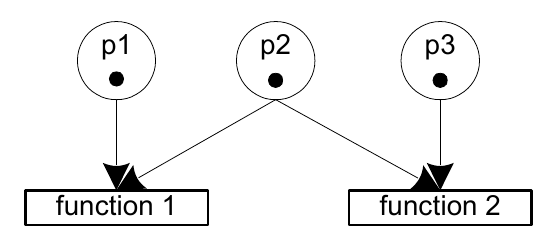}
  \end{center}
  \vspace{-10pt}
  \caption{Traditional Petri Net\newline conflict}
  \label{fig2}
\end{wrapfigure}

\subsection{Petri Net Conflict}

Figure \ref{fig2} shows the conflict pattern of ordinary Petri nets. With the 
given marking, both transitions are enabled. Firing either one 
disables the other. In EDPNs, if place p2 is replaced by a port 
input event, we have a context-sensitive input event, where outputs 
of the same physical input depend on the context in which the event 
occurs.

\begin{figure}[b]
  \vspace{-10pt}
  \centering
  \begin{minipage}[b]{0.55\textwidth}
    \centering
    \includegraphics{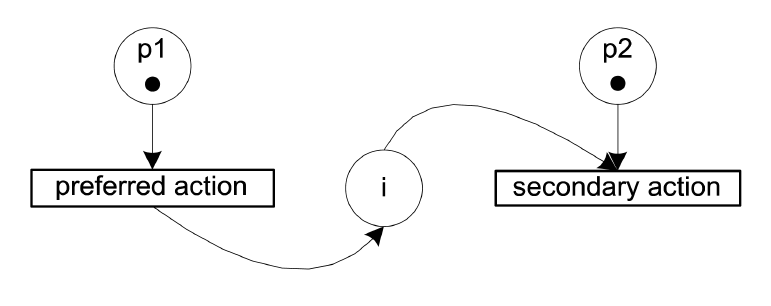}
    \vspace{-10pt}
    \caption{Traditional Petri Net\newline interlock}
    \label{fig3}
  \end{minipage}%
  \begin{minipage}[b]{0.45\textwidth}
    \centering
    \includegraphics{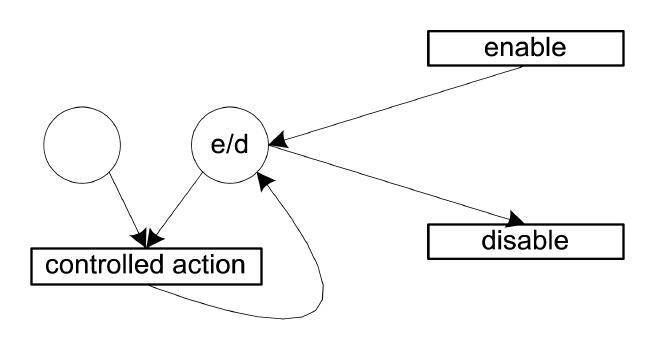}
    \vspace{-10pt}
    \caption{ESML Enable, Disable,\newline and Activate}
    \label{fig4}
  \end{minipage}
\end{figure}

\subsection{Petri Net Interlock}

The Petri net interlock pattern is used to set a priority. With the 
marking shown in Figure \ref{fig3}, the secondary action cannot fire until 
the preferred action fires. If these are linked, we have the pattern 
for mutual exclusion.

\begin{center}
  \begin{figure}
    \centering
    \includegraphics{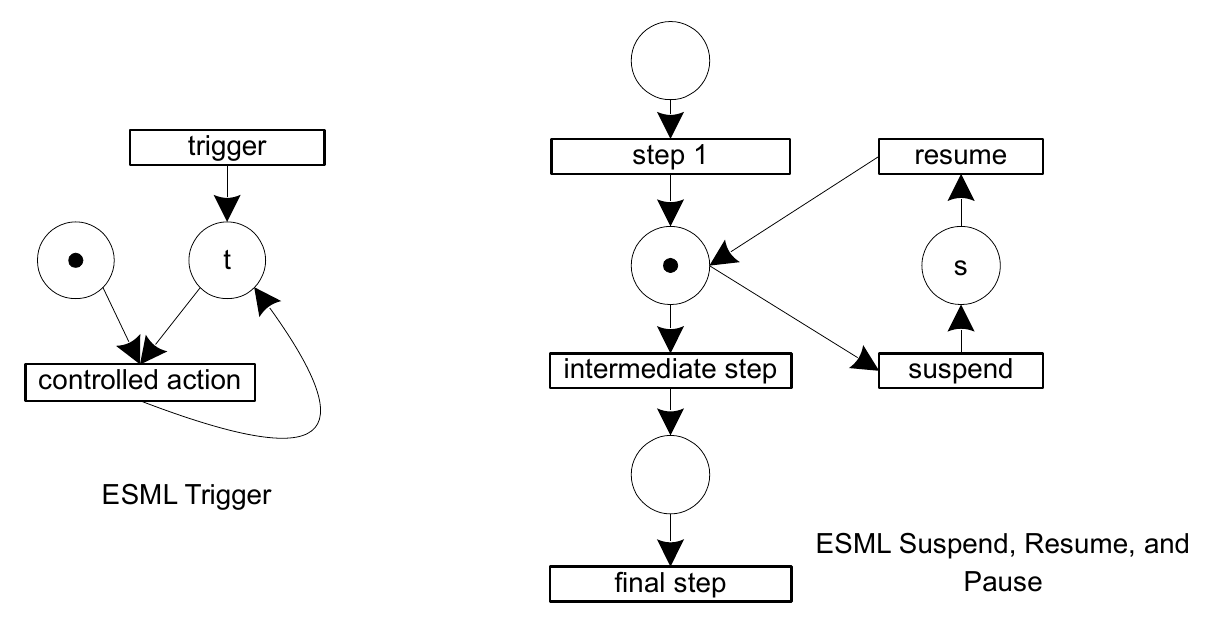}
    \caption{ESML Trigger, Suspend, Resume, and Pause}
    \label{fig5}
  \end{figure}
\end{center}

\subsection{ESML Enable, Disable, and Activate}

The work of the ESML committee began with activities in a data flow 
diagram that could control other activities. Here the place labeled 
"e/d" is used to enable, and later, to disable the controlled action 
in Figure \ref{fig4}. Clearly the enabling and disabling actions are from 
other constituents. Once a transition is enabled, it remains enabled 
due to the output leading back to the e/d place. Notice that the 
enable/disable place establishes Petri net conflict between the 
controlled and disable actions. The enable portion essentially gives 
permission for the controlled action to occur, but in terms of 
marking, it may have to await additional inputs. The edge from the 
controlled action back to the enable/disable place assures that an 
enabled transition remains enabled. The activate prompt is simply a 
sequence of enable followed by a disable.

\subsection{ESML Trigger}

The ESML trigger prompt is stronger than the enable prompt. It 
essentially requires the controlled action to fire as soon is it is 
fully enabled, as in Figure \ref{fig5}. A trigger prompt can be 
paired with a disable prompt. As with the Enable prompt, once a 
controlled action has been started by a Trigger Prompt, the Trigger 
place (marked t in Figure \ref{fig5}) remains marked due to the 
output that leads back to it. (Of course, this assumes that the 
other inputs are still available to the controlled action. As with 
the Enable prompt, a Trigger prompt can be removed by a Disable 
prompt.

\subsection{ESML Suspend, Resume, and Pause}

The ESML suspend, resume, and pause prompts (Figure \ref{fig5}) were 
designed to interact with an ongoing activity without losing any of 
the work done prior to the point of suspension. The suspend place 
``s'' is actually an interlock to assure that the resume action must 
follow a suspend action. The suspend action can be used to transfer 
temporary control to a more important action. As with activate, the 
pause prompt is a sequence of suspend followed by resume.

\subsection{Service Requests}

\begin{figure}
  \centering
  \includegraphics{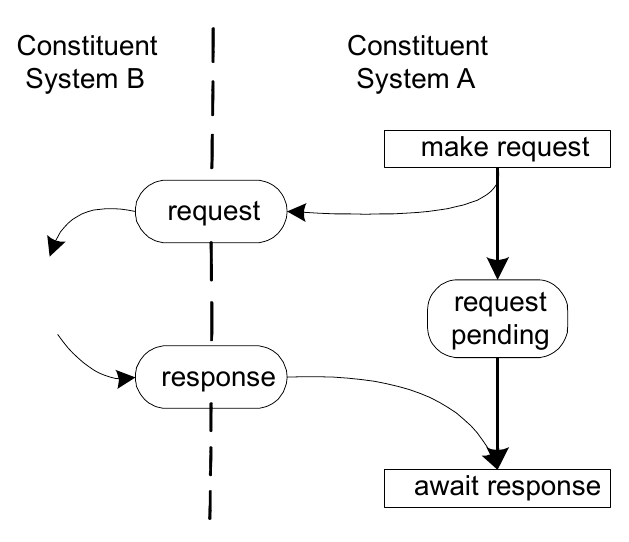}
  \caption{SoS Request}
  \label{fig6}
\end{figure}

The last three patterns are directed at communication among 
constituents. When constituent A requests a service from constituent 
B, the request place is similar to an enable prompt (Figure \ref{fig6}). 
Another interpretation is that a request is a message. Once a 
request is made, constituent A awaits a response from constituent B. 

\subsubsection{Accept and Reject Requests}

On receipt of a request, constituent B may choose to accept or 
reject the request. When the service has been provided, constituent 
B returns a "done" response to the waiting constituent A (see Figure 
8). Similarly, if constituent B rejects the request, constituent B 
returns a "not done" response to the waiting constituent. Notice 
that the request place is in Petri net conflict with respect to the 
two responses.

\begin{figure}[h]
  \centering
  \includegraphics{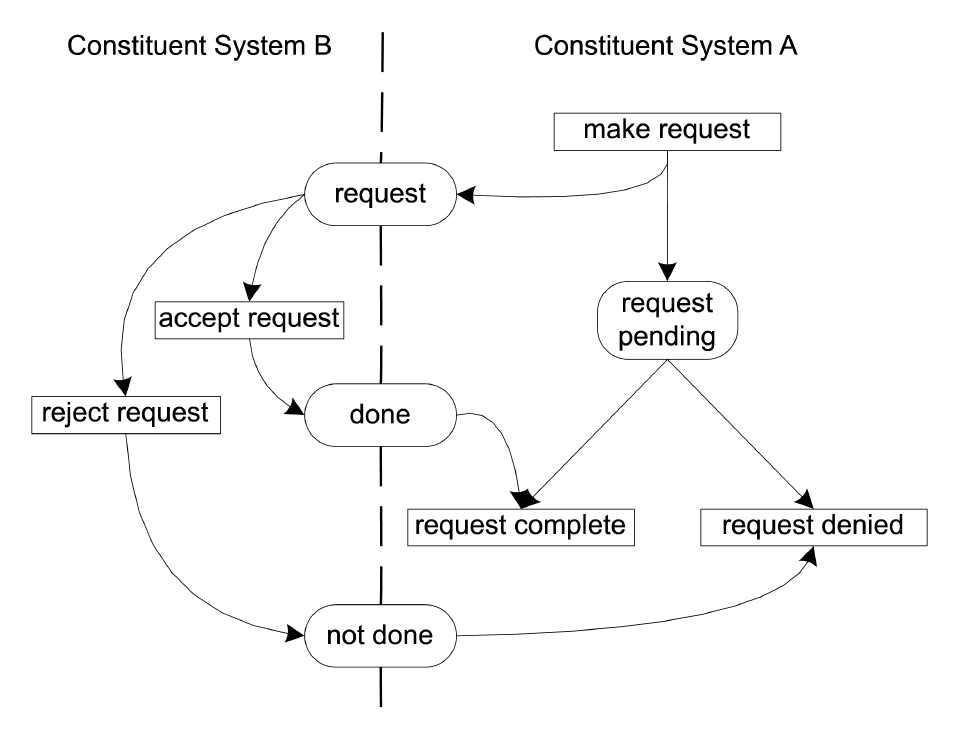}
  \caption{SoS Accept and Reject responses to a Request}
  \label{fig7}
\end{figure}

\subsubsection{Postpone Request}

\begin{figure}[h]
  \centering
  \includegraphics{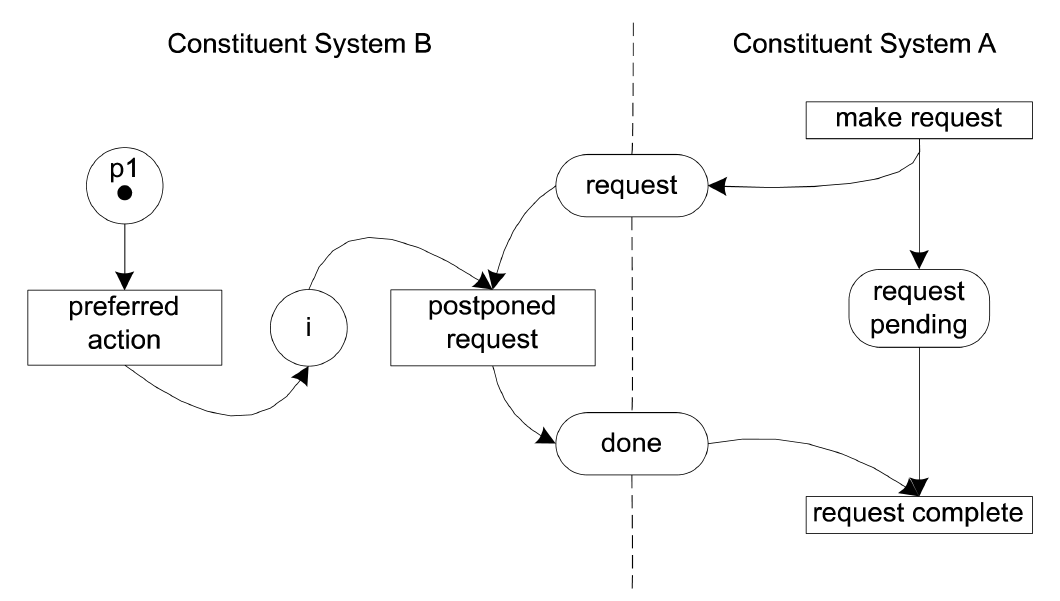}
  \caption{SoS Postpone}
  \label{fig8}
\end{figure}

Consider situations in which a constituent may have the latitude to 
postpone a request. This most likely happens because the constituent 
has more urgent tasks. Note the use of an interlock to show the task 
priority in Figure \ref{fig8}.

\section{Example: a Garage Door Controller}

\begin{figure}[h]
  \centering
  \includegraphics{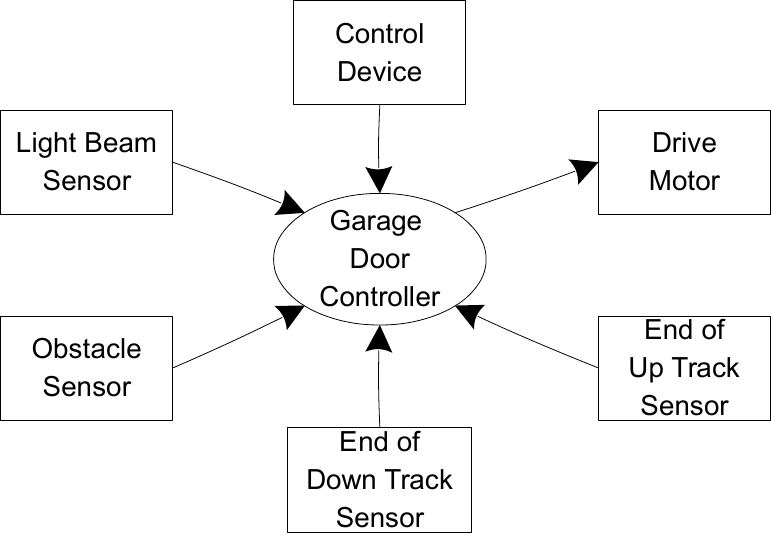}
  \caption{SysML Context Diagram of the Garage Door Controller}
  \label{fig9}
\end{figure}

A system to control a motorized garage door is comprised of several 
components: a drive motor, the garage door wheel tracks, and a 
wireless control device. There are two safety features, a light beam 
near the floor, and an obstacle sensor. These latter two devices 
operate only when the garage door is closing. If the light beam is 
interrupted (possibly by a pet) the door immediately stops, and then 
begins to open. Similarly, if the door encounters an obstacle while 
it is closing (say a child’s bicycle left in the path of the door), 
the door stops and reverses direction as with a light beam 
interruption. There is a third way to stop a door in motion, either 
when it is closing or opening—a signal from the wireless control 
device. In response to this signal the door stops in place. A 
subsequent signal starts the door in the same direction as when it 
was stopped. Finally, there are sensors that detect when the door 
has moved to one of the extreme positions, either fully open or 
fully closed. Figure \ref{fig9} is a SysML context diagram of the garage 
door controller.

This example is deliberately small, yet it suffices to illustrate 
several of the previously described interactions. Here are a few 
examples:
\begin{itemize}
  \item When the door is open, a signal from the wireless keypad 
  triggers the drive motor. 
  \item When the door is in motion, either opening or closing, a 
  signal from the wireless keypad triggers the motor to stop.  
  (Note: this portion of the problem could be interpreted as an ESML 
  Pause.) 
  \item When the door is closing, the Light Beam and the Obstacle 
  Sensors are enabled. 
  \item The fully closed sensor disables the Light Beam and the 
  Obstacle Sensors. 
  \item An input from either the Light Beam or the Obstacle Sensor 
  triggers the drive motor to stop and then begin opening.
\end{itemize}

\subsection{Statechart model of the Garage Door Controller}

\begin{figure}[h!]
  \centering
  \includegraphics[width=\textwidth]{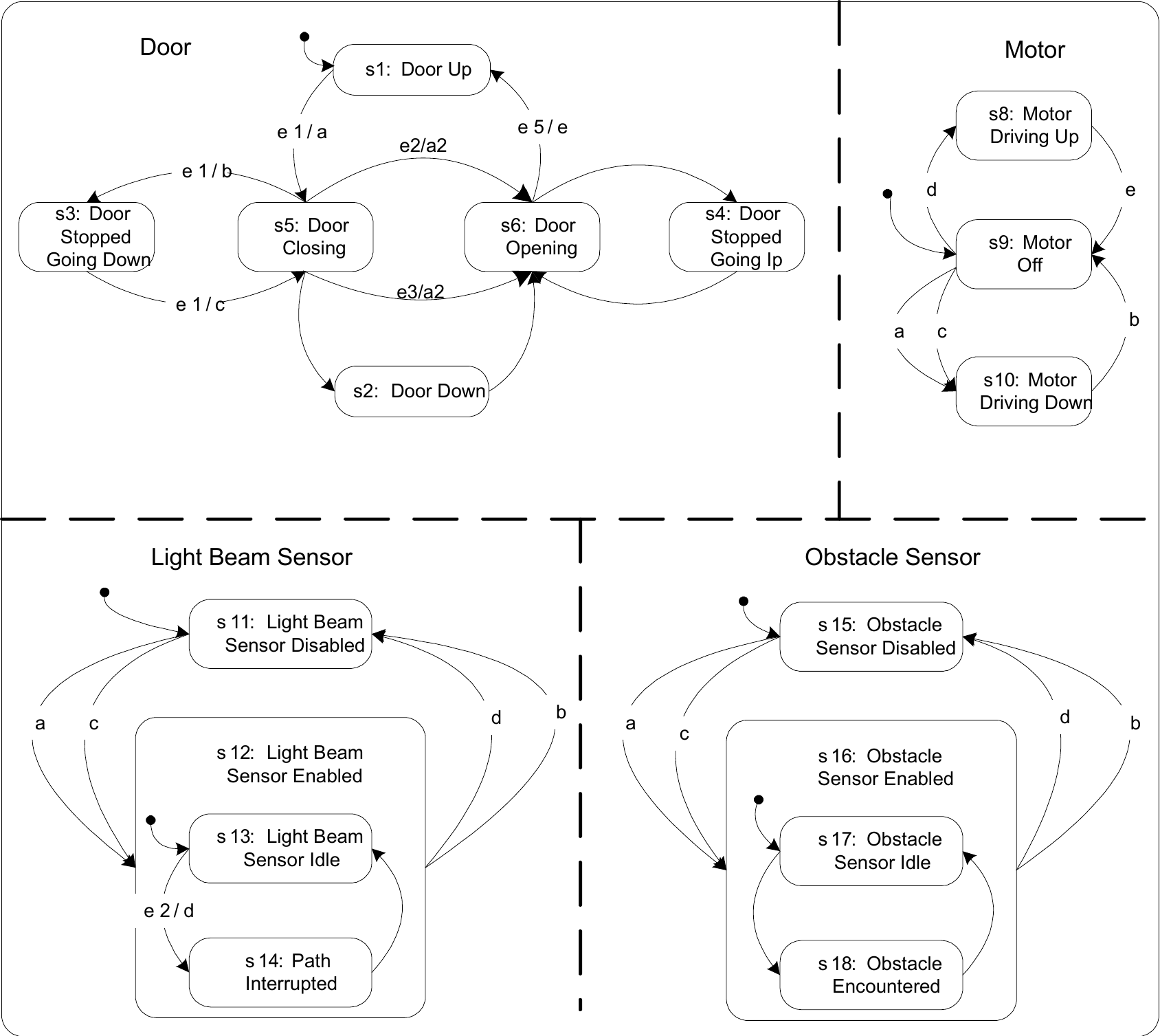}
  \caption{Full Garage Door Controller Statechart}
  \label{fig10}
\end{figure}

A full statechart model is given in Figure \ref{fig10}. The orthogonal 
regions describe local views of the full garage door, the motor, and 
both safety sensors. The input events and output actions are:\\
\begin{minipage}[t]{0.49\textwidth}
e1: wireless control signal\\
e2: light beam interruption sensed\\
e3: obstacle sensed\\
e4: end of down track reached\\
e5: end of up track reached\\
\end{minipage}
\begin{minipage}[t]{0.49\textwidth}
a1: start drive motor down\\
a2: start drive motor up\\
a3: stop drive motor\\
\end{minipage}

Some of the transitions are marked with output actions lettered a, 
b, c, d, and e. These will illustrate the broadcasting mechanism of 
statecharts.  The sequence corresponds to the Statechart execution 
for the following scenario: pre-condition: door is up, Input event 
sequence: e1, e1, e1, e2, e5, and the post-condition: door is up. 
This corresponds to what an offline model-based tester would do to 
identify a test case for the scenario. Good practice dictates making 
an "execution table," as shown in Table \ref{tab1}.

\noindent 
\begin{table}[h!]
  \caption{Execution Table for the Sample Scenario}
  \label{tab1}
  \begin{tabular}{|m{0.08\textwidth}|m{0.12\textwidth}|m{0.12\textwidth}|m{0.12\textwidth}|m{0.12\textwidth}|m{0.12\textwidth}|m{0.12\textwidth}|}
    \hline
    \multicolumn{7}{|l|}{Garage Door Statechart Execution Table} \\
    \hline
    & \multicolumn{4}{l|}{In States} & & \\
    \hline
    Step & Door & Motor & Light Beam Sensor & 
    Obstacle Sensor & Input Event & Broadcast Output \\
    \hline
    0 & s1 & s9 & s11 & s15 & e1 & a \\
    \hline
    1 & s5 & s10 & s12, s13 & s16, s17 & e1 & b \\
    \hline
    2 & s3 & s9 & s11 & s15 & e1 & c \\
    \hline
    3 & s5 & s10 & s12, s13 & s16, s17 & e2 & d \\
    \hline
    4 & s6 & s8 & s12, s14 & s15 & e5 & e \\
    \hline
    5 & s1 & s9 & s12, s14 & s15 & & \\
    \hline
  \end{tabular}
\end{table}

To use a statechart for offline MBT, the tester must begin with a 
full statechart that shows only the input events that cause 
transitions in the orthogonal regions. Next, a scenario is 
postulated, and an exercise similar to the one above is followed to 
create an execution table. This is then the skeleton of a MBT test 
case. If the tester has access to a statechart engine, the process 
is greatly simplified—the tester simply selects a starting condition 
and defines a sequence of inputs. The engine produces an execution 
table similar to Table \ref{tab1}.

\subsection{Swim Lane EDPN models of the Garage Door Controller}

\begin{figure}[h!]
  \centering
  \includegraphics{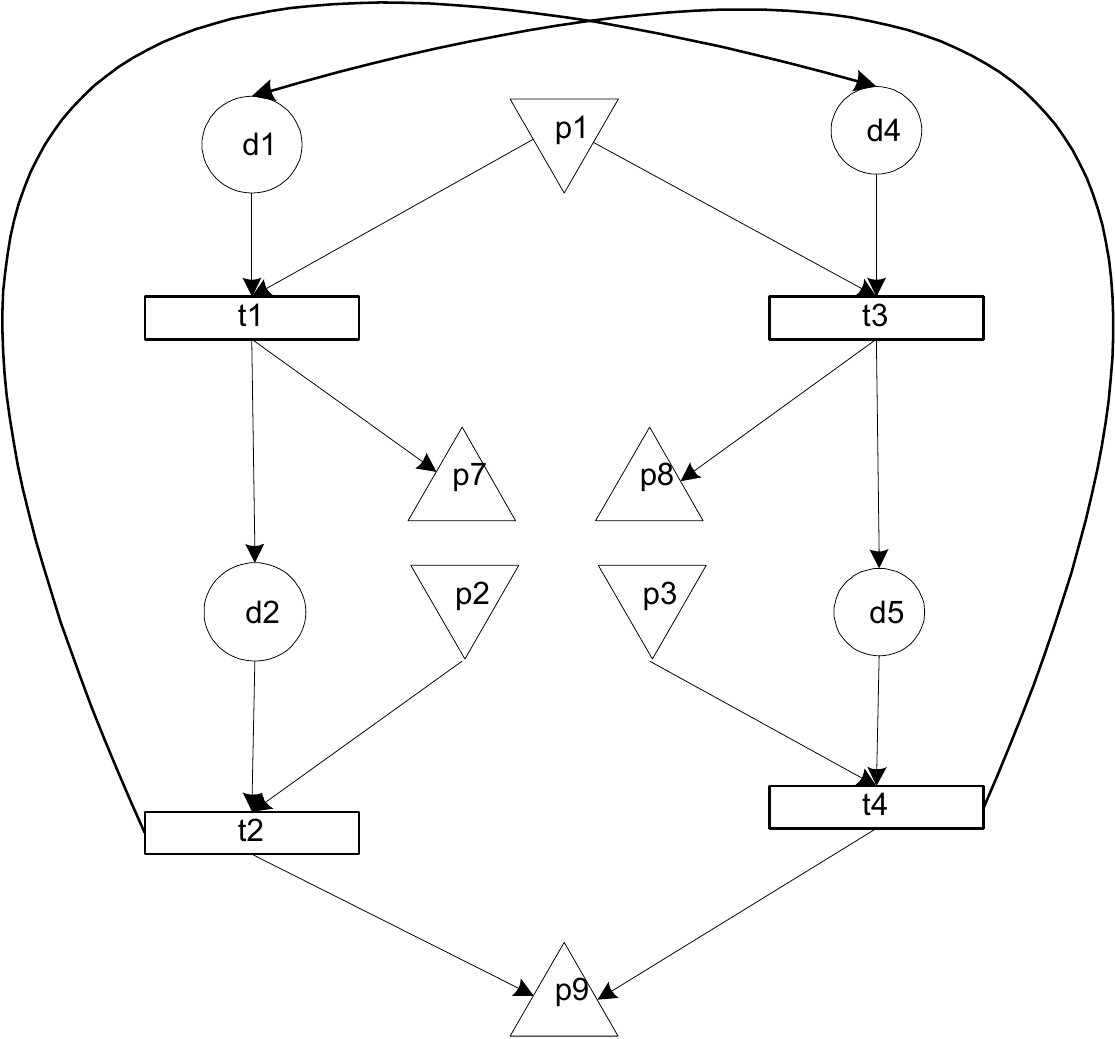}
  \caption{EDPN of the basic garage door operation}
  \label{fig11}
\end{figure}

\begin{figure}[h!]
  \centering
  \includegraphics{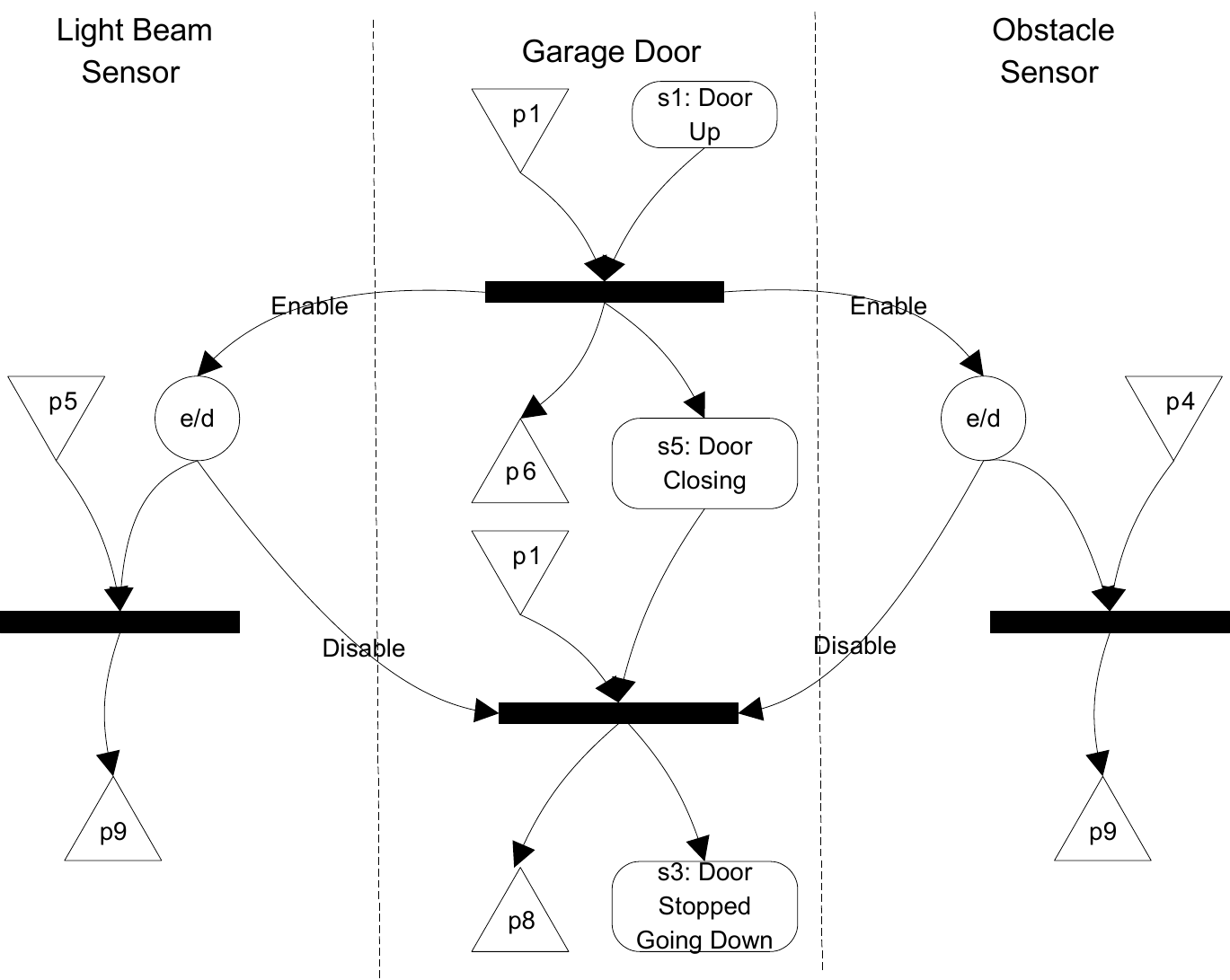}
  \caption{Swim Lane detailed view of safety feature enabling and disabling}
  \label{fig12}
\end{figure}

\begin{figure}[h!]
  \centering
  \includegraphics{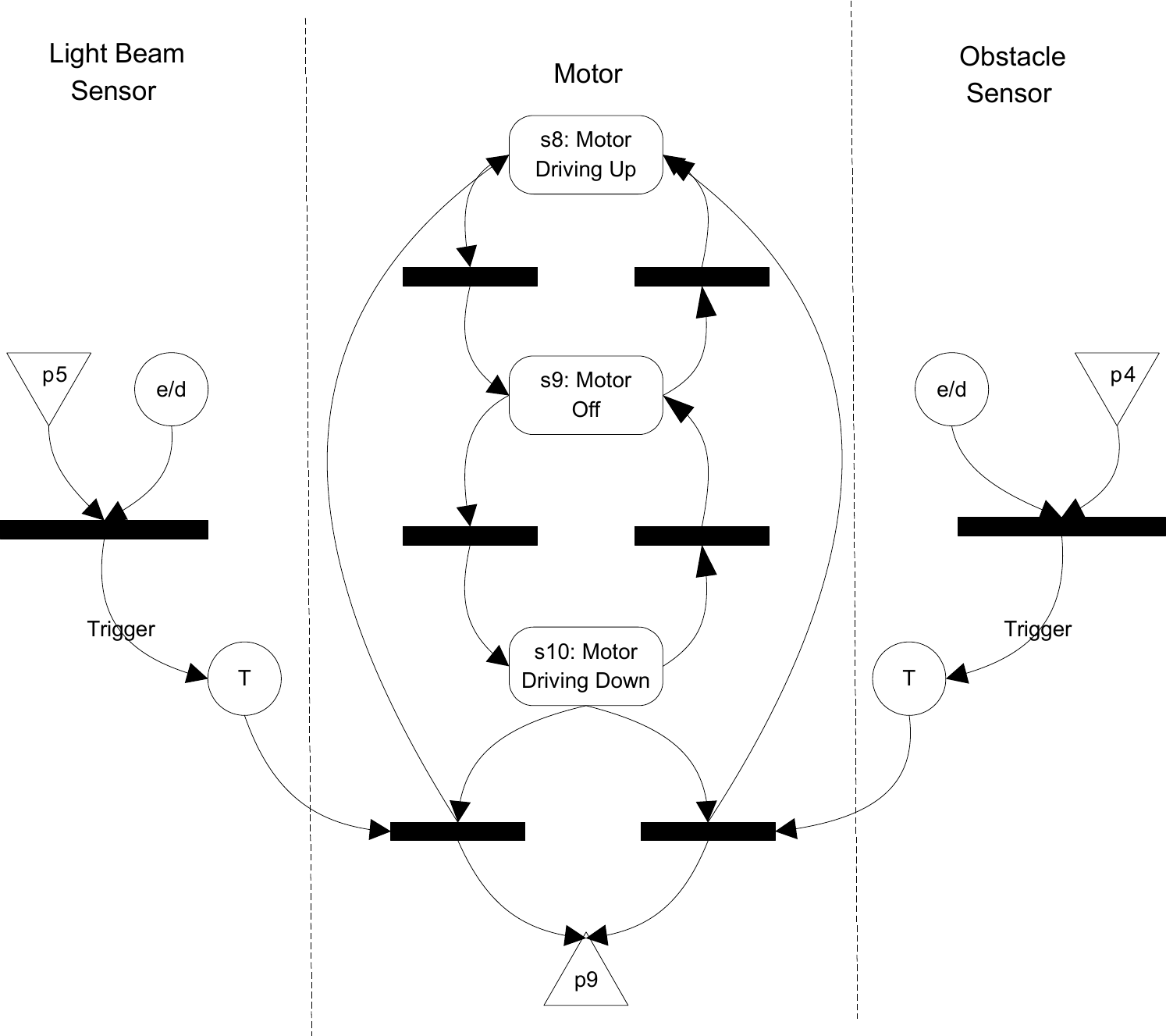}
  \caption{Swim Lane detailed view of safety feature}
  \label{fig13}
\end{figure}

\begin{figure}[h!]
  \centering
  \includegraphics[width=\textwidth]{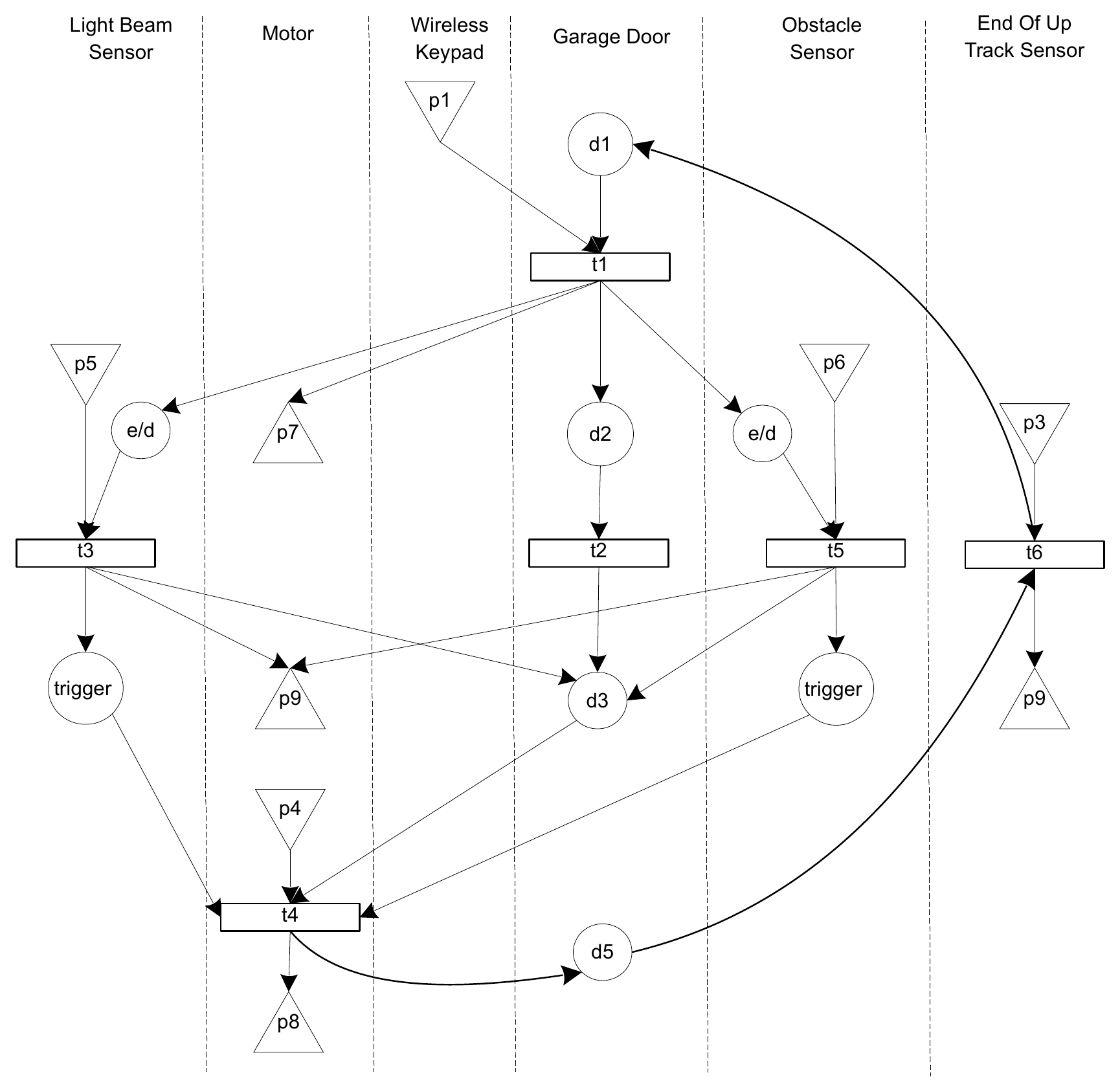}
  \caption{Swim Lanes for Light Beam and Obstacle Sensor Enabling}
  \label{fig14}
\end{figure}

Table \ref{tab2} contains a legend for all the EDPN elements in Figures \ref
{fig11}, \ref{fig12}, \ref{fig13}, \ref{fig14}, and \ref{fig16}. The 
basic operation of the Garage Door Controller is shown as an 
Event-Driven Petri Net in Figure \ref{fig12} (the intermediate stopping and 
the safety devices are omitted).  Places d1: Door Up and d4: Door 
Down are contexts for input event p1: wireless keypad signal. Since 
d1 and d4 are mutually exclusive, the context sensitivity is 
resolved. For now, assume the door is fully open (d1 is marked). 
When the transition t1 fires, output event p7: start drive motor 
down occurs, and the door is in the state d2: Door Closing. After 
some time interval (13 seconds in my garage) the door reaches the 
end of the down track, which is represented here as the input event 
p2. When transition t2 fires, the output p9: stop drive motor 
occurs, and the Garage Door Controller is in the fully closed state 
d4. If event p1 occurs again, transition t3 can fire, which causes 
output event p8 to occur and leaves the garage door in the state d5: 
Door Opening. Once the end of the up track is reached, input event 
p3 occurs, transition t4 fires, the motor is stopped (output event 
p9), and the door is back in the fully open state (d1). When 
described as an Event-Driven Petri Net (without swim lanes), we have 
the full picture, but we miss the interactions among devices. The 
interactions are apparent, and we could show the overall execution 
by a marking sequence.

\noindent
\begin{table}
  \caption{Swim Lane EDPN elements for the Garage Door Controller}
  \label{tab2}
  \centering
  \begin{tabular}{|p{0.3\textwidth}|p{0.3\textwidth}|p{0.3\textwidth}|}
    \hline
    Input events & Output events (actions) & Data Places \\
    \hline
    p1:  wireless keypad signal & p7: start drive motor down & d1: Door 
    Up \\
    \hline
    p2:  end of down track hit & p8:  start drive motor up & d2: Door 
    Closing \\
    \hline
    p3:  end of up track hit & p9:  stop drive motor & d3: Door Stopped 
    going down \\
    \hline
    p4:  brief motor pause &  & d4: Door Down \\
    \hline
    p5:  light beam sensor &  & d5: Door Opening \\
    \hline
    p6: obstacle sensor &  & d6: Door Stopped going up \\ 
    \hline
  \end{tabular}
\end{table}

System test cases can be derived directly from an Event-Driven Petri 
Net. A system test case corresponds to a sequence of transitions 
that fire. Since the EDPN in Figure \ref{fig11} is 3-connected (a true path 
exists to and from every transition) there can be a countable 
infinite set of distinct paths. Table \ref{tab3} lists three sample paths.

\noindent
\begin{table}
  \caption{Selected Paths in Figure \ref{fig10}}
  \label{tab3}
  \centering
  \begin{tabular}{|p{0.4\textwidth}|p{0.25\textwidth}|}
    \hline
      Path Description & Transition Sequence \\
    \hline
      1.  Close an open garage door. & t1, t2 \\
    \hline
      2.  Open a closed garage door. & t3, t4 \\
    \hline
      3.  Open closed door and then close it. & t3, t4, t1, t2 \\
    \hline
  \end{tabular}
\end{table}

Deriving a full system test case is straightforward. The test case 
for Path 1 in Table \ref{tab2} is:

\noindent \textit{Name}: 				Close an open garage door.\\
\textit{Pre-conditions}:		Garage Door is open

\noindent \textit{Event Sequence}\\
\begin{tabular}{p{0.4\textwidth}l}
  Input Events	&	Output Events \\
  1.  p1: wireless keypad signal	&	2.  p7: start drive motor down\\
  3.  p2:  end of down track hit	&	4.  p9:  stop drive motor\\
\end{tabular}

\noindent \textit{Post-conditions}: Garage Door is closed
  
Figure \ref{fig12} shows a detailed view of the interaction between the 
garage door closing and the enabling/disabling of the two safety 
features. This view is more appropriate for offline model-based 
testing, as it allows the tester to focus on specific interactions. 
In Figure \ref{fig13}, the interactions between the motor and the two safety 
devices are shown. This can be considered to be a continuation of 
the enabling and disabling of the safety features shown in Figure 
13. If either safety device input occurs, the Trigger prompt 
immediately causes the motor to reverse and drive the garage door to 
the open position. Notice that Figures 13 and 14 could be composed 
into a larger, more expressive, Swim Lane EDPN. This becomes 
cumbersome, and even unwieldy quickly, as we see in Figure \ref{fig14}.

For comparison, look at Figure \ref{fig14}~-- a fairly complete Swim Lane 
Event-Driven Petri Net showing the enabling of the safety features 
(light beam and obstacle sensor) and how an event from either device 
stops and then reverses the garage door motion. In Figure \ref{fig14}, assume 
an initial marking of place d1: Door Up. If event p1: wireless 
keypad signal occurs, transition t1 fires with four results: the 
light beam sensor and the obstacle sensors are enabled, the drive 
motor is started in the down (closing) direction, and the Garage 
Door is in the d2: Door Closing state. If event p5:  light beam 
sensor occurs, transition t3 fires, causing output event p9:  stop 
drive motor, leaving the garage door in state d3: Door Stopped going 
down. (The scenario for the obstacle sensor is symmetric to that for 
the light beam sensor.) The next event is p4:  brief motor pause, 
which allows transition t4 to fire because it has been triggered, 
and this causes output event p8:  start drive motor up to occur, 
leaving the garage door in state d5: Door Opening. Event p3 occurs 
when the end of the up track is reached, which stops the drive motor 
(p9) leaving the garage door in state d1: Door Up.

\section{Concluding Thoughts}

Swim Lane Event Driven Petri Nets provide several advantages for 
model-based testing:
\begin{itemize}
  \item they are a bottom-up approach
  \item they can be easily composed
  \item they permit focused description of interactions among constituents
  \item they can be used for automatic derivation of system test cases
  \item they support a useful hierarchy of system test coverage metrics.
\end{itemize}

\begin{figure}[h!]
  \centering
  \includegraphics{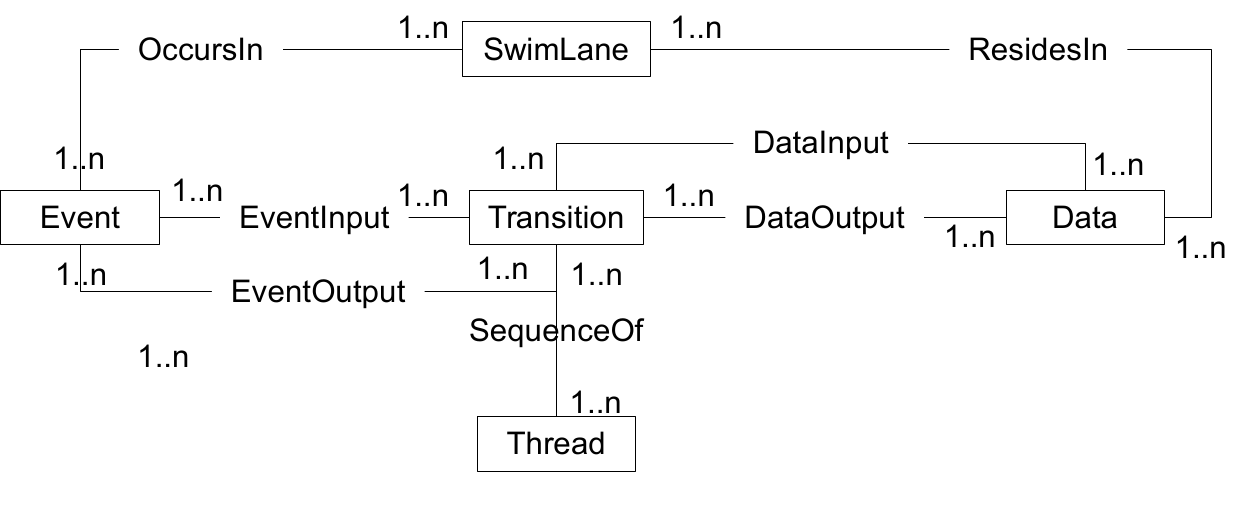}
  \caption{E/R Model of Swim Lane EDPN Database}
  \label{fig15}
\end{figure}

The biggest limitation of Swim Lane Event Driven Petri Nets is that 
the drawings do not scale up well. They are vulnerable to a 
diagrammatic explosion similar to the "finite state machine 
explosion." All (graphical) Petri net models suffer from the problem 
of space-consuming diagrams. Figure \ref{fig15} presents an elegant answer to 
this issue. Rather than compose Swim Lane EDPN diagrams, we can 
populate a database with the E/R description in Figure \ref{fig15}. With such 
a formulation, questions of connectivity are reduced to 
well-constructed database queries. One clear advantage of this is 
that now there is no practical limit to Swim Lane EDPN composition.

\subsection{Graphical Composition of Event-Driven Petri Nets}

\begin{figure}[h!]
  \centering
  \includegraphics{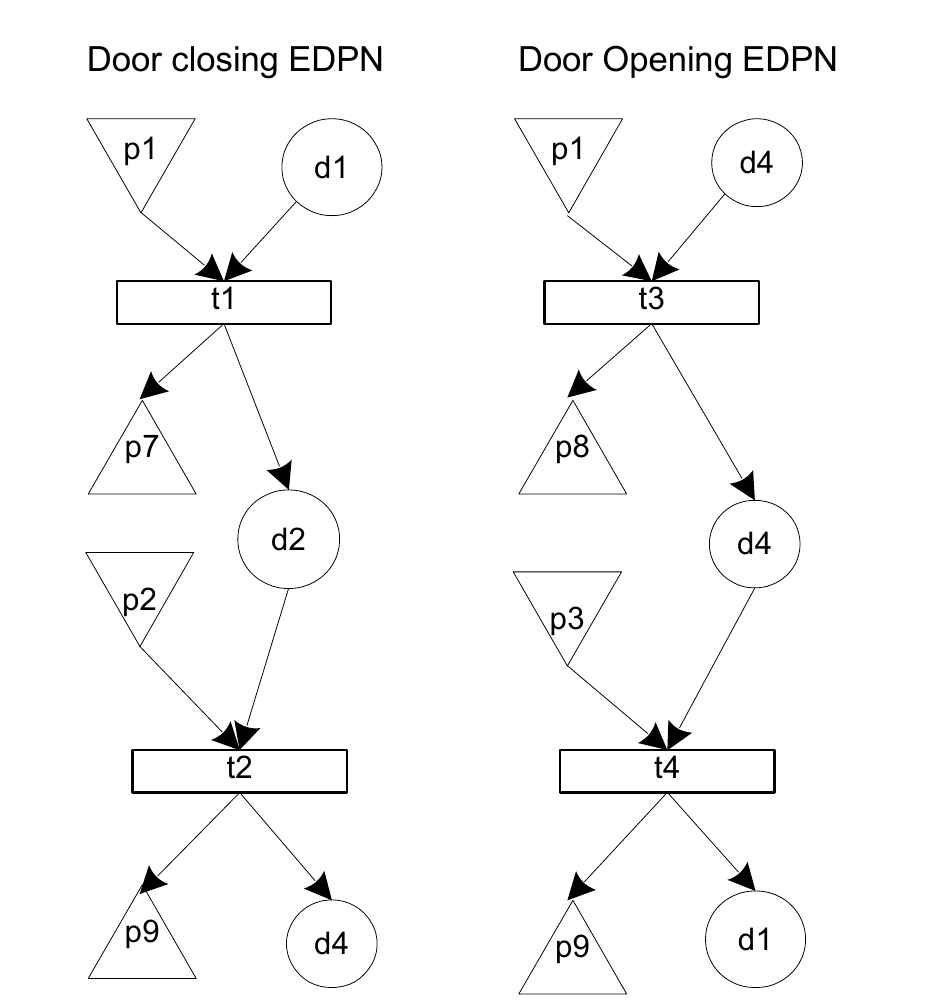}
  \caption{Individual Door Closing and Door Opening Event-Driven Petri Net}
  \label{fig16}
\end{figure}

Graphical composition of Event-Driven Petri Nets is straightforward, 
but it easily expands to spatial difficulty. Figure \ref{fig16} shows two 
Event-Driven Petri Nets for closing and opening the garage door; 
their composition was shown earlier in Figure \ref{fig11}.

\subsection{Composition of Event-Driven Petri Nets in the E/R Model 
Database}

\noindent \begin{tabular}{|p{0.075\textwidth}|p{0.10\textwidth}||p{0.075\textwidth}|p{0.10\textwidth}||p{0.075\textwidth}|p{0.10\textwidth}||p{0.075\textwidth}|p{0.10\textwidth}|}
    \hline
    \multicolumn{8}{|l|}{Database for Door Closing EDPN} \\
    \hline
      \multicolumn{2}{|l||}{EventInput} &
      \multicolumn{2}{l||}{EventOutput} &
      \multicolumn{2}{l||}{DataInput} &
      \multicolumn{2}{l|}{DataOutput} \\
    \hline
      Event& Transition & Event & Transition & Data & Transition &
        Data & Transition \\
    \hline
      p1 & t1 & P7 & t1 & d1 & t1 & d2 & t1 \\
    \hline
      p2 & t2 & P9 & t2 & d2 & t2 & D4 & t2 \\
    \hline
\end{tabular}

\noindent \begin{tabular}{|p{0.075\textwidth}|p{0.10\textwidth}||p{0.075\textwidth}|p{0.10\textwidth}||p{0.075\textwidth}|p{0.10\textwidth}||p{0.075\textwidth}|p{0.10\textwidth}|}
    \hline
    \multicolumn{8}{|l|}{Database for Door Opening EDPN} \\
    \hline
      \multicolumn{2}{|l||}{EventInput} &
      \multicolumn{2}{l||}{EventOutput} &
      \multicolumn{2}{l||}{DataInput} &
      \multicolumn{2}{l|}{DataOutput} \\
    \hline
      Event& Transition & Event & Transition & Data & Transition &
        Data & Transition \\
    \hline
      p1 & t3 & p7 & t3 & d3 & t3 & d4 & t3 \\
   \hline
      p3 & t4 & p8 & t4 & d4 & t4 & d1 & t4 \\
   \hline
\end{tabular}

\noindent \begin{tabular}{|p{0.075\textwidth}|p{0.10\textwidth}||p{0.075\textwidth}|p{0.10\textwidth}||p{0.075\textwidth}|p{0.10\textwidth}||p{0.075\textwidth}|p{0.10\textwidth}|}
    \hline
    \multicolumn{8}{|l|}{Database for Composition of Door Closing and 
    Door Opening EDPNs} \\
    \hline
      \multicolumn{2}{|l||}{EventInput} &
      \multicolumn{2}{l||}{EventOutput} &
      \multicolumn{2}{l||}{DataInput} &
      \multicolumn{2}{l|}{DataOutput} \\
    \hline
      Event& Transition & Event & Transition & Data & Transition &
        Data & Transition \\
    \hline
      p1 & t1 & p6 & t1 & d1 & t1 & d2 & t1 \\
    \hline
      p1 & t3 & p7 & t3 & d2 & t2 & d3 & t2 \\
    \hline
      p2 & t2 & p8 & t2 & d3 & t3 & d4 & t3 \\
    \hline
      p3 & t4 & p8 & t4 & d4 & t4 & d1 & t4 \\
   \hline
\end{tabular}

\subsection{Further Guidelines}

As a guideline, Swim Lane EDPNs are best used to focus on particular 
interactions, and then using the database approach to keep the 
overall model. While test cases can be derived by inspection of a 
graphical model, this is more difficult from the underlying 
database. Finally, Swim Lane EDPNs support the definition of the set 
of system level test coverage metrics given in Table \ref{tab4}. These 
metrics all refer to a set of test cases T derived from a 
corresponding Swim Lane EDPN description.

\begin{table}[h!]
  \caption{Test Coverage Metrics}
  \label{tab4}
  \centering
  \begin{tabular}{|p{0.15\textwidth}|l|}
    \hline
    Test Cover & Description \\
    \hline
    Ct & every transition\\
    \hline
    Cp & every data place \\
    \hline
    Cie & every input event \\
    \hline
    Coe & every output event \\
    \hline
    Ccontext & Cie in every context \\
    \hline
  \end{tabular}
\end{table}

\nocite{*}
\bibliographystyle{eptcs}
\bibliography{generic}
\end{document}